\documentclass[showpacs,apl,twocolumn,floatfix]{revtex4}

\usepackage{graphicx,float,amsmath,mathrsfs,amssymb}

\begin{document}

\title{The effect of Pauli blockade on spin-dependent diffusion in a degenerate electron gas}
\author{F. Cadiz}
\email{fabian.cadiz@polytechnique.edu}
\affiliation{Physique de la mati\`ere condens\'ee, Ecole Polytechnique, CNRS, 91128
Palaiseau, France}
\author{D. Paget}
\affiliation{Physique de la mati\`ere condens\'ee, Ecole Polytechnique, CNRS, 91128
Palaiseau, France}
\author{A.C.H. Rowe}
\affiliation{Physique de la mati\`ere condens\'ee, Ecole Polytechnique, CNRS, 91128
Palaiseau, France}

\begin{abstract}
Spin-polarized transport of photo-electrons in bulk, p-type GaAs is investigated in the Pauli blockade regime. In contrast to usual spin diffusion processes in which the spin polarization decreases with distance traveled due to spin relaxation, images of the polarized photo-luminescence reveal a spin-filter effect in which the spin polarization \textit{increases} during transport over the first 2 $\mu$m from 26 \% to 38 \%. This is shown to be a direct consequence of the Pauli Principle and the associated quantum degeneracy pressure which results in a spin-dependent increase in the minority carrier diffusion constants and mobilities. The central role played by the quantum degeneracy pressure is confirmed via the observation of a spin-dependent increase in the photo-electron volume and a spin-charge coupling description of this is presented.

\end{abstract}

\maketitle

Spin-polarized transport in semiconductors has been studied intensively for more than a decade. Compared to metals, semiconductors are of interest because spin lifetimes and diffusion lengths are long \cite{appelbaum2007,kikkawa1999}, and because the orientation of electron spins can be controlled electrically \cite{koga2002}. These two characteristics are necessary for a large number of proposed active, spintronic devices \cite{awschalom2007}. As such, any physical interaction which modifies spin transport is not only of scientific interest but also of potential practical importance. Recently a number of theoretical predictions concerning novel spin-transport phenomena have been made by modifying the spin diffusion equations to include various possible coupling effects \cite{vignale2000,takahashi2008}. One particular coupling - the Coulomb spin drag effect - has been explicitly dealt with and was recently observed experimentally \cite{weber2005}. While implicitly included in these descriptions, one element whose direct consequences on spin-polarized transport are yet to be explored is the Pauli Principle.

The Pauli Principle is a key element in the quantum mechanical description of nature \cite{cohen1977}. Its basic premise, that two Fermions may not simultaneously occupy the same quantum state, has profound consequences for a number of apparently disparate physical systems \cite{shapiro1983,pauling1960}. At high Fermion densities (in the so-called Pauli blockade regime) it predicts the appearance of a quantum degeneracy pressure that was recently observed under controlled conditions in an atom trap \cite{truscott2001}, and that manifests itself as an increase of the charge carrier diffusion constant and mobility in solids \cite{smith1978}. Here it is shown that the quantum degeneracy pressure has at least two major consequences for spin-polarized transport in semiconductors. The first is the appearance of a spin-filter effect in which the spin-polarization of quantum degenerate electrons increases with the distance over which they are transported. Even in the presence of other novel spin-dependent phenomena \cite{weber2005, weber2009}, this observation is contrary to the usual expectations of spin-polarized transport in which the spin-polarization decreases with distance due to spin relaxation \cite{appelbaum2007,kikkawa1999}. Secondly, a spin-charge coupling appears that causes an increase the volume of the photo-electron population at high electronic polarizations. This increase in volume is equivalent to that observed in degenerate atom traps \cite{truscott2001}.

\begin{figure}[tbp]
\includegraphics[clip,width=8 cm] {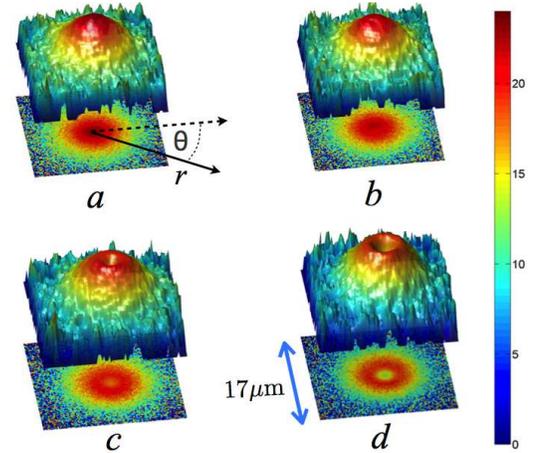}
\caption{Pauli blockade driven spin-filter effect.
Images of the luminescence polarization, $\mathscr{P} \times |\mathscr{P}_i|$, in a GaAs thin film at $T = 15$ K induced by a circularly polarized, tightly focused laser spot at $r=0$. The images correspond to excitation powers of a) $72$ nW, b) $0.45$ mW, c) $1.89$ mW, d) $2.55$ mW. As the excitation power and thus the photo-electron concentration increases, the appearance of a polarization dip at $r=0$  is observed i.e. the polarization increases during outwards diffusion. Luminescence spectra indicate an electronic temperature, $T_e=50$ K, near $r=0$.
}
\label{fig1}
\end{figure}

In the experiments reported here, a spin-polarized degenerate photoelectron gas is created in a p-type GaAs ($N_A=10^{18} \; \mbox{cm}^{-3}$) thin film of thickness $d = 3 \; \mu$m by a continuous circularly-polarized (left, $\sigma^{-}$ or right, $\sigma^{+}$) laser beam of wavelength $780$ nm. The beam is focused to a diffraction-limited Gaussian spot of half-width $\omega= 0.6 \; \mu$m. Under these conditions, the photon angular momentum is partially transferred to the photo-electrons and the initial electronic (spin) polarization is $\mathscr{P}_i= \mp 50 \; \%$ for $\sigma^{\pm}$ polarized light \cite{meier1984}. A polarized micro-luminescence experiment \cite{favorskiy2010} with a spatial resolution of $0.2 \; \mu $m is used to image the spatial distribution of photo-electron charge, $n(r)=n_{+}+n_{-}$, the spin $s(r)=n_{+}-n_{-}$, and the spin polarization, $\mathscr{P}=s/n$ at temperatures between $6$ K and $300$ K. Here $n_{-}$ ($n_{+}$) are the concentrations of electrons with spin aligned parallel (antiparallel) to the direction of light propagation. The variations of $n$ and $s$ on depth do not depend strongly on radial position coordinate $r$. As such, for a qualitative approach, the diffusion in this sample is effectively 2-dimensional.   

Charge and spin transport were first characterized at $T = 15$ K at low excitation power. In this case the charge concentration $n(r)$ decreases from $r=0$ over a characteristic diffusion length $L=\sqrt{D_0 \tau}$ where $D_0$ is the spin-independent diffusion constant and $\tau$ is the minority carrier lifetime \cite{favorskiy2010}. The spin distribution decays over a different, characteristic spin diffusion length $L_s=\sqrt{D_s \tau_s}$ where $\tau_s$ is the spin lifetime. Since the spin lifetime $\tau_s \leqslant \tau$, $\mathscr{P}(r)$ also drops from its maximum value at $r=0$ towards zero as $r \rightarrow \infty$. This case is shown in Fig. \ref{fig1}(a), where the luminescence polarization is $23 \; \%$ at the center (implying a spin polarization $\mathscr{P}(0)=46 \; \%$) before dropping to immeasurably small values for $r > 10 \; \mu$m. By analyzing the luminescence images \cite{favorskiy2010}, the characteristic diffusion lengths are found to be $L = 1.45$ $\mu$m and $L_s = 1.23$ $\mu$m for the charge and spin respectively. A separate, time-resolved photoluminescence experiment was used to determine $\tau=270 \pm 50$ ps and $\tau_s=220 \pm 50$ (data not shown). Combining these times with the measured diffusion lengths yields the diffusion constants $D_0 \approx 82 \pm 10 \;\mbox{cm}^2/s$ and $D_s \approx 71 \pm 15 \;\mbox{cm}^2/s$, so that within experimental uncertainty $D_0 = D_s$ and spin-drag \cite{weber2005} is absent. The surface recombination velocity ($S = 4.6 \times 10^{4}$ cm/s) is almost a factor of 10 smaller than the diffusion velocity, $D_0/d \approx 2.73 \times 10^{5}$ cm/s, so that surface recombination is negligible. These parameters are used below to numerically model the effects of Pauli blockade effects with no adjustable parameters.

\begin{figure}[tbp]
\includegraphics[clip,width=8 cm] {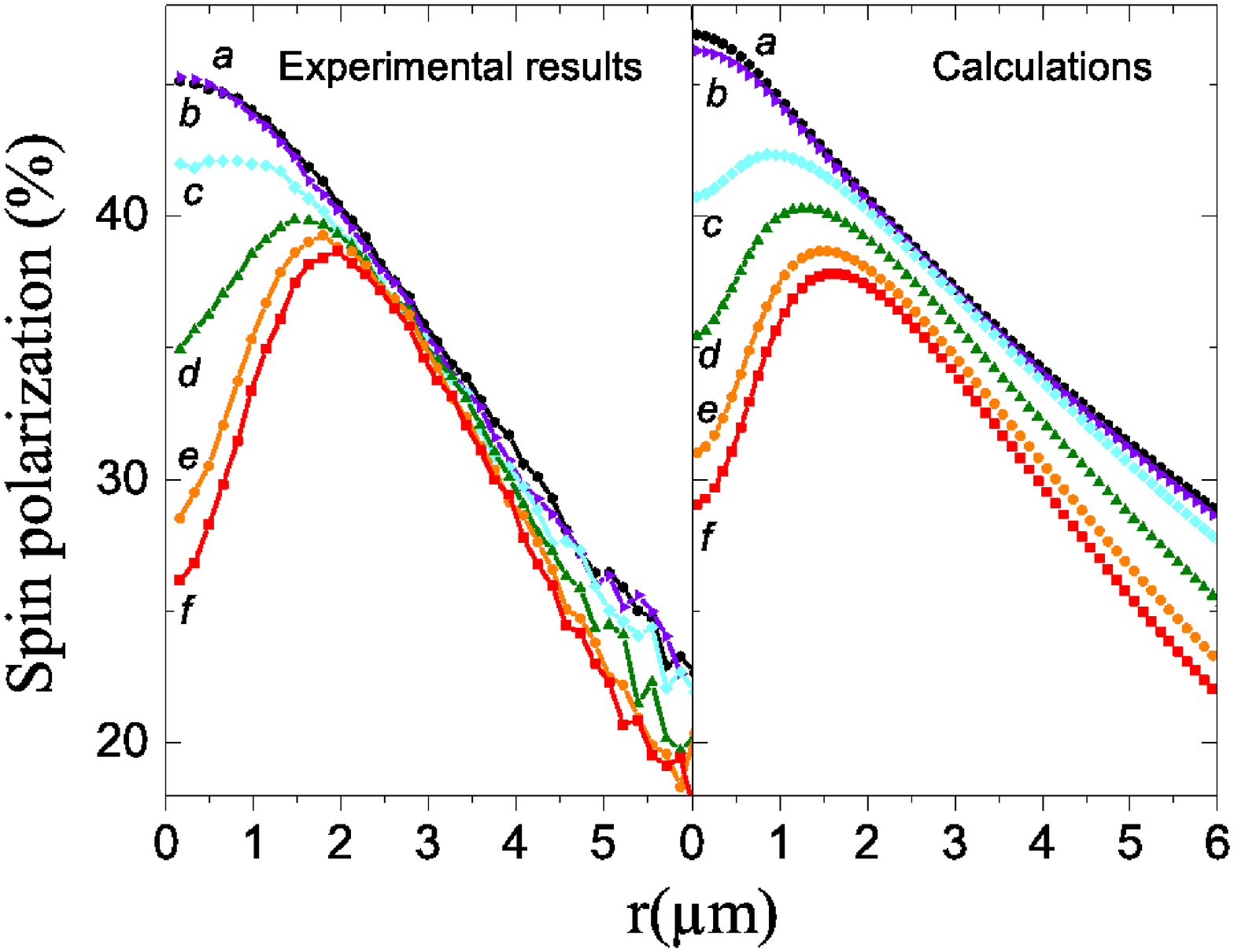}
\caption{Experimental results and comparison with the model. 
Left panel, the measured spin polarization as a function of space and excitation power for a) $0.0072$ mW, b) $0.028$ mW, c) $0.45$ mW, d) $1.03$ mW, e) $1.89$ mW, and f) $2.55$ mW. The sample is kept at $T = 15$ K, resulting in an average electronic temperature of $T_e=50$ K. The right panel shows a numerical resolution of Eqs.(\ref{eq2}) and (\ref{eq3}). The excellent agreement between the data and the model, obtained with no adjustable parameters, indicates that Coulomb spin-drag is of secondary importance. 
}
\label{fig2}
\end{figure}

As the excitation power is progressively increased, $n(0)$ becomes comparable to and eventually larger than the conduction band effective density of states, $N_c=4.3 \times 10^{17} (T_e/300)^{3/2}\; \mbox{cm}^{-3}$ where $T_e$ is the electronic temperature. Figure \ref{fig1}(b), Fig. \ref{fig1}(c) and Fig. \ref{fig1}(d) show that under these conditions the luminescence polarization (and therefore $\mathscr{P}(r)$) exhibit a dip near $r=0$ and counter-intuitively increase during diffusive transport for $0 < r < 2 \; \mu m$. This observation is not only in complete opposition to usual notions of spin relaxation and diffusion \cite{appelbaum2007,kikkawa1999,favorskiy2010}, but cannot be described by any of the other novel, spin transport phenomena that have recently been reported \cite{weber2005,weber2009}. For the following discussion, the angular ($\theta$) averaged spin polarization profiles corresponding to the images in Fig.  \ref{fig1} are shown in Fig. \ref{fig2}, and the value of $\mathscr{P}(0)$ can be read off these curves.

\begin{figure}[tbp]
\includegraphics[clip,width=8 cm] {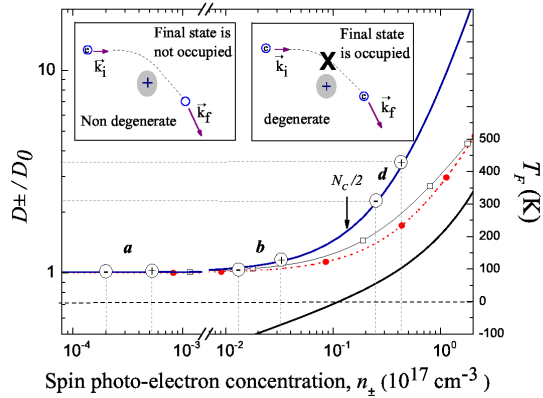}
\caption{Reduced photo-electron diffusion constant. 
$D_\pm/D_0=\nu $ [defined in Eq. (\ref{eq1})] at $T_e= 50$ K for GaAs as a function of  $n_\pm$, the majority and minority spin concentrations. Also shown are the concentration dependence of the reduced mobility $\zeta(n_\pm)$ (filled, red circles), of the ratio $\xi(n_\pm)$ of the diffusion constant to the mobility (black, open squares), and of the Fermi temperature of the electron gas. The calculated spin-concentrations $n_\pm$ corresponding to panels a, b, and d are shown (panel c has been omitted for clarity). Pauli blockade manifests itself as a concentration and spin dependence of the diffusion constant at high concentration. 
}
\label{fig3}
\end{figure}

As will be shown here, this effect is a direct consequence of the Pauli Principle and, more specifically, of the quantum degeneracy pressure that yields a dependence of the diffusion constant on the spin concentrations $n_{\pm}$, as summarized in Fig. \ref{fig3}. It is assumed a priori that, after the initial excitation, the majority and minority spin populations thermalize to quasi-equilibria associated with two quasi-Fermi energies, $E_{F_\pm}$, with $E_{F_+} > E_{F_-}$. At low excitation powers, such as in the left side of Fig. \ref{fig3}, both quasi-Fermi levels fall in the bandgap of the semi-conductor and the concentrations $n_+$ and $n_-$ are both smaller than $N_c/2$. In this limit and in the absence of Coulomb spin drag \citep{weber2005}, the diffusion constant for both spin populations is given by the Einstein relation, $D_0=\mu_0  k_B T_e/q$ where $k_B$ is Boltzmann's constant, $q$ is the absolute value of the electron charge and $\mu_0$ is the electron mobility. In the extreme opposite case both $E_{F_+}$ and $E_{F_-}$ lie above the bottom of the conduction band and $n_+$ and $n_-$ are larger than $N_c/2$. In this degenerate limit, the Pauli Principle dictates a reduction in the charge carrier scattering rates via final-state blocking (see the inset schematic diagrams in Fig. \ref{fig3}) and consequently an increase in the average momentum relaxation time, $\tau_m$. This results in an increase in the mobility \cite{flatte2006}, given (in the likely case of scattering by ionised impurities) by \begin{equation} \label{mobility} \mu_\pm=\mu_0 (\sqrt{\pi}/4) F_2 (E_{F_\pm} )/F_{1/2} (E_{F_\pm} )=\mu_0 \zeta(n_\pm ), \end{equation} and also in an increase of $\xi(n_\pm )=D(n_\pm)/\mu(n_\pm)$, both of which are functions of $\tau_m$ and hence $n_\pm$.  Here $F_j$ is the Fermi integral of index $j$.  The general expression for the diffusion constant is \cite{smith1978} \begin{equation} D_\pm= \frac{\mu_0 k_B T_e}{q}  \sqrt{\pi}/2  F_2 (E_{F_\pm} )/F_{1/2} (E_{F_\pm})=D_0 \nu(n_\pm) \label{eq1} \end{equation} where $\nu(n_\pm)= \xi(n_\pm )\zeta(n_\pm)$. Eq. (\ref{eq1}) is strictly equivalent to the (spin) susceptibility formulation of the generalised Einstein relationship \cite{finkelshtein1983}. Shown in Fig. \ref{fig3} are the variations of $\zeta$, $\xi$, and $\nu$ as a function of $n_\pm$. The function $\nu$ can differ significantly from 1 for $n_\pm > N_c/2$, and in the strongly degenerate case ($n_\pm \gg N_c/2$), the high energy expansion of the Fermi integral yields $D_\pm \propto n_{\pm}^{5/3}$ \cite{sommerfeld1928}. Thus, since $n_+ > n_-$, $D_+ > D_-$ and majority spin electrons diffuse further than minority spin electrons. Consequently there is an effective depletion of majority carriers in the central region of the images in Fig.  \ref{eq1}, bringing  $n_+$ closer to  $n_-$ and thereby reducing the luminescence polarization relative to its value at low excitation powers. Therefore, a dip in polarization is expected near $r= 0$. As $r$ further increases, both $n_+$ and $n_-$ are reduced via carrier recombination until they are smaller than $N_c/2$. In this case neither spin population is degenerate and the luminescence polarization begins to decrease with increasing $r$ as usual.\

A semi-quantitative analysis of the experimental results is now performed by estimating the spin concentrations $n_\pm$ in a small region of surface area $\omega^2$ centered at $r = 0$. In this case, spin relaxation and photo-electron recombination can be considered negligible compared to outward diffusion processes i.e. $\tau,\tau_s \gg  \omega^2/(4D_\pm )$. The effective spin lifetimes are therefore $\omega^2/(4D_\pm )$ (within numerical factors of order unity \cite{paget2012}). For continuous photo-excitation the concentrations at $r=0$ are given by the non-linear equation: $n_\pm \approx g_\pm(0) \omega^2/[4 D_0 \nu(n_\pm)]$, where Eq. (\ref{eq1}) has been used \cite{favorskiy2010}. Here $g_\pm(0)$ are the spin dependent generation rates of photo-electrons (at $r = 0$) per unit volume. Their ratio in GaAs photo-excited at $780$ nm is $g_+(0)/g_-(0)=(1+\mathscr{P}_i^{*})/(1-\mathscr{P}_i^{*})$. Here, in order to account for losses by spin relaxation during thermalisation \cite{meier1984}, the effective value of $\mathscr{P}_i^{*}=0.45$ is taken equal to $\mathscr{P}(0)$ at low power rather than to $\mathscr{P}_i= 0.5$ expected from the optical selection rules . Figure \ref{fig3} shows the values of $n_\pm$  calculated for panels a, b, and d of Fig.  \ref{fig1} (panel c has been omitted for clarity) using a numerical resolution of the above non-linear equation at $T_e=50$ K. While panel a and, to some extent, b correspond to the non-degenerate case, in the case of panel d, $n_+ \approx 4.6 \times 10^{16}\; \mbox{cm}^{-3}$ and $n_- \approx 2.7 \times 10^{16}\; \mbox{cm}^{-3}$, both larger than $N_c/2$. Using Eq. (\ref{eq1}) this gives  $D_+= 3.4 \;D_0$ and $D_-=2.3 \; D_0$  implying a relative difference of $45\; \%$ between the two diffusion constants. The mobility is also found to depend significantly on spin with $\mu_+/\mu_- = 1.2$. Note also that the characteristic excitation power, \textbf{P}, required for the onset of the Pauli blockade effect is that which yields $n(0) \approx N_c$. One finds \textbf{P} $\approx 1$ mW, in good agreement with the transition observed between curves c and d of Fig.  \ref{fig2}.

\begin{figure}[tbp]
\includegraphics[clip,width=8 cm] {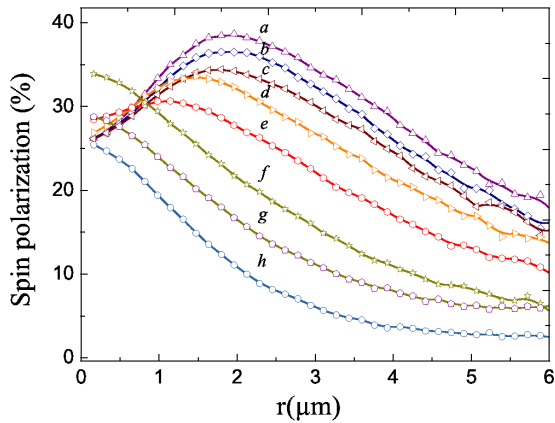}
\caption{Temperature dependence of spin-dependent diffusion. 
Spin polarization as a function of space at excitation power of $2.55$ mW and different values of the electronic temperature $T_e$ , a) $55$ K, b) $57$ K, c) $68$ K, d) $82$ K, e) $90$ K, f) $110$ K, g) $200$ K and h) $360$ K. The transition from the Pauli blockade to the non-degenerate regime occurs between $90$ K and $110$ K, in agreement with the expected temperature variation of the effective density of states, $N_c$.
}
\label{fig4}
\end{figure}

An increase of electron temperature will result in an increase of the effective density of states, proportional to $T_{e}^{3/2}$, thus reducing the degree of degeneracy and the effects of Pauli blockade. Figure \ref{fig4}, which shows the spin polarization profiles obtained for different electronic temperatures at an excitation power of \textbf{P} $= 2.55$ mW, indeed reveals a transition from a degenerate to a non-degenerate gas between  $T_e=90$ K (curve e) and $T_e =110$ K (curve f). The electronic temperature that determines the transition into the non-degenerate regime, obtained by writing $n(0)=N_c$, and by neglecting the weak temperature dependence of $D_0$, is $110$ K. This is in excellent agreement with the temperature evolution of the polarization profiles of Fig.  \ref{fig4}.\\

A quantitative interpretation of the data for $r \neq 0$ requires a  numerical solution of the coupled spin and charge diffusion equations. If Coulomb spin-drag  is neglected \cite{vignale2000,weber2005,takahashi2008}, then to first order in $\eta \mathscr{P}$, \begin{equation} \label{eta} \eta=\frac{n/2}{\nu(n/2)}\frac{d\nu(n/2)}{dn} \end{equation} and the spin-dependent diffusion constant from Eq. (\ref{eq1}) may be written $D_\pm = D(1 \pm \eta \mathscr{P})$ where $D=D_0 \nu(n/2)$. Under these conditions, the coupled diffusion equations have a simple form \begin{equation} g-n/\tau+ \vec \nabla \cdot [D(\vec \nabla n + \eta \mathscr{P}\vec \nabla s)]=0 \label{eq2} \end{equation} \begin{equation} \mathscr{P}_i g- n/\tau_s + \vec \nabla \cdot [D(\vec \nabla s+ \eta \mathscr{P}\vec \nabla n)]=0 \label{eq3} \end{equation} where $g=g_+(r)+ g_-(r)$ is the electron-hole pair creation rate. More details on the calculation of luminescence and polarization profiles can be found in the Supplementary Information.

Two effects play a marginal role in the quantitative determination of $n(r)$ and $s(r)$. Firstly, ambipolar diffusion cannot be completely ignored, in which case $D$ must be replaced by the ambipolar diffusion coefficient $D_a$ \cite{smith1978,paget2012}. Since $D_a$ is smallest near $r=0$, $n$ is increased at the center thereby augmenting the effects of Pauli blockade. Secondly, it is necessary to account for possible modifications of spin-lattice relaxation at high power due to Pauli blockade \cite{amo2007}  or to a local increase of $T_e$ by about $50$ K, \cite{zerrouati1988}  as found from the shape of the luminescence spectrum. The change of spin relaxation time is revealed in the decrease of the spatially-averaged electronic spin polarization, from $39\; \%$ to $32\; \%$ with increasing excitation power (data not shown). By assuming that this reduction is due entirely to a reduction in the spatially-independent value of $\tau_s$, the variation of $\tau_s$ with excitation power can be estimated and empirically included in the numerical resolution of Eqs. (\ref{eq2}) and (\ref{eq3}).\

The calculation spin polarization profiles at $T= 15$ K are shown in the right panel of Fig. \ref{fig2} next to the angular averaged profiles taken from the images in Fig.  \ref{fig1}. The agreement between experiment and calculation is excellent and confirms that the spin-filter effect in which $\mathscr{P}$ increases during transport is primarily due to Pauli blockade. \

The relative increase in the majority spin diffusion constant in the Pauli blockade regime is physically equivalent to saying that the majority spins experience an additional outwards force due to quantum degeneracy pressure. Using the values of $n_+$ and $n_-$ obtained at $T= 15$ K and $r= 0$, along with the equation of state for a Fermion gas at $T= 0$ K, $(p_\pm= 2n_\pm k_B T_{F_\pm})/5$, the highest excitation power yields a pressure, $p_+ \approx 670$ Pa ($p_- \approx 250$ Pa) compared with the classical result $p_+=nk_B T_e \approx 460$ Pa. Here the Fermi temperatures ($T_{F_\pm}$) can be read off the right hand axis of Fig.  \ref{fig3} (black curve). Thus the quantum degeneracy pressure for the majority spin population is a factor of $1.5$ times higher than the classical result whereas for minority spins the contribution is comparable to the classical pressure.\

\begin{figure}[tbp]
\includegraphics[clip,width=8 cm] {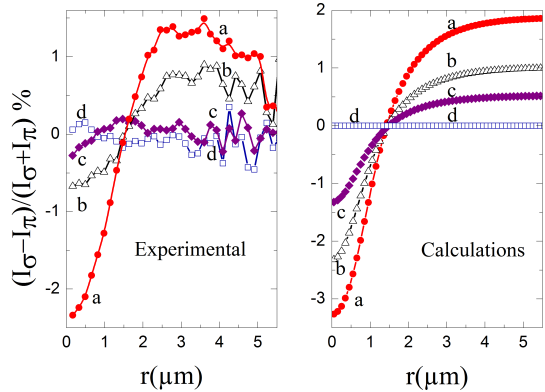}
\caption{Charge-spin coupling driven by Pauli blockade.
Left panel, relative difference between the luminescence intensity profiles obtained under circularly polarized excitation $\sigma$ ($I_\sigma$) and under linearly-polarized excitation $\pi$ ($I_\pi$), for a power of a) $2.33$ mW, b) $0.95$ mW, c) $0.41$ mw and d) $65$ nW. For each excitation power, the only difference between $\sigma$ and $\pi$ excitation is the initial spin polarization of the photo-electrons. A difference of the order of $-2.5 \;\%$ between the profiles is observed at high power at $r=0$, in excellent agreement with the diffusion constant variations shown in Fig. \ref{fig3}. Right panel, the angular integrated, theoretical prediction given by a numerical solution of Eq. (\ref{eq2}) and Eq. (\ref{eq3}). 
}
\label{fig5}
\end{figure}

This quantum degeneracy pressure should also change the volume of the spin-polarized electron gas in the Pauli blockade regime in analogy with observations in atomic systems \cite{truscott2001}. It is possible to directly observe this phenomenon by modulating the photo-excitation polarization between a $\sigma^+$  circularly polarized state and a linear (or $\pi$) polarized state without changing the excitation power. In the latter case $\mathscr{P}_i = 0 \;\%$ so the photo-electrons are no longer spin polarized. On the other hand, as long as the excitation power is not modified, the photo-electron creation rate is identical for both $\sigma^+$ and $\pi$ excitations. Figure \ref{fig5} shows the relative variation in the photo-electron density profiles due to this change in light polarization at a lattice temperature of $15$ K. Qualitatively one notes that for progressively higher photo-excitation powers, if the photo-electron gas is polarized then $n$ decreases near $r=0$ relative to the un-polarized case. polarization of the electron gas pushes the majority spins further into the Pauli blockade regime and yields a spin-averaged photo-electron population that does indeed occupy a larger volume in space. This extra majority spin pressure is at the origin of both the spin-filter effect described above as well as the observed increase in the spin-averaged photo-electron volume.\

The data in the left panel of Fig. \ref{fig5} can be modelled (as above) by obtaining numerical solutions to Eqs.(\ref{eq2}) and (\ref{eq3}). The results of this calculation are shown in the right panel of Fig. \ref{fig5}. Again it is possible to make an order-of-magnitude estimate of the effect of the spin degeneracy pressure at $r=0$. It can be shown that the mean electronic diffusion constant, defined as $\langle D \rangle=(n_+ D_{+} + n_- D_-)/n$  is given by \begin{equation} \label{diffusion} \langle D \rangle=D(1+\eta \mathscr{P}^2), \end{equation} where $\eta$ is defined in Eq. \ref{eta}. In the case of $\pi$-excitation, $\mathscr{P} = 0\; \%$ and $\langle D \rangle =D$. A simple estimate of the relative photo-electron concentration difference at $r= 0$ is then $(n_\sigma -n_\pi)/(n_\sigma +n_\pi)=-(\langle D_\pi \rangle - \langle D_\sigma \rangle)/(\langle D_\sigma \rangle + \langle D_\pi \rangle)=- \mathscr{P}(0)^2 \eta /(2+\eta \mathscr{P}(0)^2)$. For $P = 2.55$ mW, $(n_\sigma-n_\pi)/(n_\sigma+n_\pi )\approx -2.7\; \%$ which is very close to the difference of measured luminescence intensities observed in Fig.  \ref{fig5} (left panel) at high excitation powers. The value of $\eta \approx 0.8$ obtained at high power indicates that the electron gas is not in the strongly degenerate regime where, according to the above discussion of the concentration dependence of the diffusion constant it should be equal to $5/3$ (see Fig.  \ref{fig3}).

In summary, in a spin-polarized electron gas with, at minimum, a majority spin density bigger than $N_c/2$, quantum degeneracy induces a spin filter effect in which both the mobility and diffusion constant depend on spin. This is well described as a coupling between the charge and spin diffusion equations. While this letter deals with bulk GaAs at low temperature, it would be of interest to explore the effect of Pauli blockade in lower dimensional systems since quantum confinement can increase $|\mathscr{P}_i |$ to $100\; \%$ while $N_c$ is smaller and less sensitive to changes in temperature. In this case, the effects of Pauli blockade should persist to higher temperatures, opening the way to their exploitation in room temperature semiconductor spintronic devices.

\section{Acknowledgements}
The authors thank S. Arscott and E. Peytavit for providing the GaAs sample, P. Barate for help with the sample characterization using time-resolved photo-luminescence techniques, and J.-P. Korb for help with the use of Laplace transform techniques. One of us (F. C.) is grateful to CONICYT Grant Becas Chile for supporting his work.

\bibliographystyle{apsrev}

\bibliography{cadizref}

\begin{thebibliography}{21}
\expandafter\ifx\csname natexlab\endcsname\relax\def\natexlab#1{#1}\fi
\expandafter\ifx\csname bibnamefont\endcsname\relax
  \def\bibnamefont#1{#1}\fi
\expandafter\ifx\csname bibfnamefont\endcsname\relax
  \def\bibfnamefont#1{#1}\fi
\expandafter\ifx\csname citenamefont\endcsname\relax
  \def\citenamefont#1{#1}\fi
\expandafter\ifx\csname url\endcsname\relax
  \def\url#1{\texttt{#1}}\fi
\expandafter\ifx\csname urlprefix\endcsname\relax\def\urlprefix{URL }\fi
\providecommand{\bibinfo}[2]{#2}
\providecommand{\eprint}[2][]{\url{#2}}

\bibitem[{\citenamefont{Appelbaum et~al.}(2007)\citenamefont{Appelbaum, Huang,
  and Monsma}}]{appelbaum2007}
\bibinfo{author}{\bibfnamefont{I.}~\bibnamefont{Appelbaum}},
  \bibinfo{author}{\bibfnamefont{B.}~\bibnamefont{Huang}}, \bibnamefont{and}
  \bibinfo{author}{\bibfnamefont{D.~J.} \bibnamefont{Monsma}},
  \bibinfo{journal}{Nature} \textbf{\bibinfo{volume}{447}},
  \bibinfo{pages}{295} (\bibinfo{year}{2007}).

\bibitem[{\citenamefont{Kikkawa and Awschalom}(1999)}]{kikkawa1999}
\bibinfo{author}{\bibfnamefont{J.~M.} \bibnamefont{Kikkawa}} \bibnamefont{and}
  \bibinfo{author}{\bibfnamefont{D.~D.} \bibnamefont{Awschalom}},
  \bibinfo{journal}{Nature} \textbf{\bibinfo{volume}{397}},
  \bibinfo{pages}{4853} (\bibinfo{year}{1999}).

\bibitem[{\citenamefont{Koga et~al.}(2002)\citenamefont{Koga, Nitta, Akazaki,
  and Takayanagi}}]{koga2002}
\bibinfo{author}{\bibfnamefont{T.}~\bibnamefont{Koga}},
  \bibinfo{author}{\bibfnamefont{J.}~\bibnamefont{Nitta}},
  \bibinfo{author}{\bibfnamefont{T.}~\bibnamefont{Akazaki}}, \bibnamefont{and}
  \bibinfo{author}{\bibfnamefont{H.}~\bibnamefont{Takayanagi}},
  \bibinfo{journal}{Phys. Rev. Lett} \textbf{\bibinfo{volume}{89}},
  \bibinfo{pages}{046801} (\bibinfo{year}{2002}).

\bibitem[{\citenamefont{Awschalom and Flatte}(2007)}]{awschalom2007}
\bibinfo{author}{\bibfnamefont{D.~D.} \bibnamefont{Awschalom}}
  \bibnamefont{and} \bibinfo{author}{\bibfnamefont{M.~E.}
  \bibnamefont{Flatte}}, \bibinfo{journal}{Nature Phys}
  \textbf{\bibinfo{volume}{3}}, \bibinfo{pages}{153} (\bibinfo{year}{2007}).

\bibitem[{\citenamefont{D'Amico and Vignale}(2000)}]{vignale2000}
\bibinfo{author}{\bibfnamefont{I.}~\bibnamefont{D'Amico}} \bibnamefont{and}
  \bibinfo{author}{\bibfnamefont{G.}~\bibnamefont{Vignale}},
  \bibinfo{journal}{Phys. Rev. B} \textbf{\bibinfo{volume}{62}},
  \bibinfo{pages}{4853} (\bibinfo{year}{2000}).

\bibitem[{\citenamefont{Takahashi et~al.}(2008)\citenamefont{Takahashi, Inaba,
  and Hiroshe}}]{takahashi2008}
\bibinfo{author}{\bibfnamefont{Y.}~\bibnamefont{Takahashi}},
  \bibinfo{author}{\bibfnamefont{N.}~\bibnamefont{Inaba}}, \bibnamefont{and}
  \bibinfo{author}{\bibfnamefont{F.}~\bibnamefont{Hiroshe}},
  \bibinfo{journal}{J. Appl. Phys.} \textbf{\bibinfo{volume}{104}},
  \bibinfo{pages}{023714} (\bibinfo{year}{2008}).

\bibitem[{\citenamefont{Weber et~al.}(2005)\citenamefont{Weber, Gedik, Moore,
  J.Orenstein, Stephens, and Awschalom}}]{weber2005}
\bibinfo{author}{\bibfnamefont{C.~P.} \bibnamefont{Weber}},
  \bibinfo{author}{\bibfnamefont{N.}~\bibnamefont{Gedik}},
  \bibinfo{author}{\bibfnamefont{J.~E.} \bibnamefont{Moore}},
  \bibinfo{author}{\bibnamefont{J.Orenstein}},
  \bibinfo{author}{\bibfnamefont{J.}~\bibnamefont{Stephens}}, \bibnamefont{and}
  \bibinfo{author}{\bibfnamefont{D.~D.} \bibnamefont{Awschalom}},
  \bibinfo{journal}{Nature} \textbf{\bibinfo{volume}{437}},
  \bibinfo{pages}{1330} (\bibinfo{year}{2005}).

\bibitem[{\citenamefont{Cohen-Tannoudji}(1977)}]{cohen1977}
\bibinfo{author}{\bibfnamefont{C.}~\bibnamefont{Cohen-Tannoudji}},
  \emph{\bibinfo{title}{Quantum Mechanics I \& II}} (\bibinfo{publisher}{Wiley,
  New York}, \bibinfo{year}{1977}).

\bibitem[{\citenamefont{Shapiro and Teukolsky}(1983)}]{shapiro1983}
\bibinfo{author}{\bibfnamefont{S.~L.} \bibnamefont{Shapiro}} \bibnamefont{and}
  \bibinfo{author}{\bibfnamefont{S.~A.} \bibnamefont{Teukolsky}},
  \emph{\bibinfo{title}{Black Holes, White Dwarfs and Neutron Stars, The
  Physics of Compact Objects}} (\bibinfo{publisher}{Wiley, New York},
  \bibinfo{year}{1983}).

\bibitem[{\citenamefont{Pauling}(1960)}]{pauling1960}
\bibinfo{author}{\bibfnamefont{L.}~\bibnamefont{Pauling}},
  \emph{\bibinfo{title}{The Nature of the Chemical Bond}}
  (\bibinfo{publisher}{Cornell University Press, New York},
  \bibinfo{year}{1960}).

\bibitem[{\citenamefont{Truscott et~al.}(2001)\citenamefont{Truscott, Strecker,
  McAlexander, Partridge, and Hulet}}]{truscott2001}
\bibinfo{author}{\bibfnamefont{A.~G.} \bibnamefont{Truscott}},
  \bibinfo{author}{\bibfnamefont{K.~E.} \bibnamefont{Strecker}},
  \bibinfo{author}{\bibfnamefont{W.~I.} \bibnamefont{McAlexander}},
  \bibinfo{author}{\bibfnamefont{G.~B.} \bibnamefont{Partridge}},
  \bibnamefont{and} \bibinfo{author}{\bibfnamefont{R.~G.} \bibnamefont{Hulet}},
  \bibinfo{journal}{Science} \textbf{\bibinfo{volume}{291}},
  \bibinfo{pages}{2570} (\bibinfo{year}{2001}).

\bibitem[{\citenamefont{Smith}(1978)}]{smith1978}
\bibinfo{author}{\bibfnamefont{R.~A.} \bibnamefont{Smith}},
  \emph{\bibinfo{title}{Semiconductors}} (\bibinfo{publisher}{Cambridge
  University Press, Cambridge}, \bibinfo{year}{1978}).

\bibitem[{\citenamefont{Koralek et~al.}(2009)\citenamefont{Koralek, Weber,
  Orenstein, Bernevig, Zhang, Mack, and Awschalom}}]{weber2009}
\bibinfo{author}{\bibfnamefont{J.~D.} \bibnamefont{Koralek}},
  \bibinfo{author}{\bibfnamefont{C.~P.} \bibnamefont{Weber}},
  \bibinfo{author}{\bibfnamefont{J.}~\bibnamefont{Orenstein}},
  \bibinfo{author}{\bibfnamefont{B.~A.} \bibnamefont{Bernevig}},
  \bibinfo{author}{\bibfnamefont{S.~C.} \bibnamefont{Zhang}},
  \bibinfo{author}{\bibfnamefont{S.}~\bibnamefont{Mack}}, \bibnamefont{and}
  \bibinfo{author}{\bibfnamefont{D.~D.} \bibnamefont{Awschalom}},
  \bibinfo{journal}{Nature} \textbf{\bibinfo{volume}{458}},
  \bibinfo{pages}{610} (\bibinfo{year}{2009}).

\bibitem[{\citenamefont{Meier and Zakharchenya}(1984)}]{meier1984}
\bibinfo{author}{\bibfnamefont{F.}~\bibnamefont{Meier}} \bibnamefont{and}
  \bibinfo{author}{\bibfnamefont{B.}~\bibnamefont{Zakharchenya}},
  \bibinfo{journal}{Optical Orientation (North-Holland, Amsterdam)}
  (\bibinfo{year}{1984}).

\bibitem[{\citenamefont{Favorskiy et~al.}(2010)\citenamefont{Favorskiy, Vu,
  Peytavit, Arscott, Paget, and Rowe}}]{favorskiy2010}
\bibinfo{author}{\bibfnamefont{I.}~\bibnamefont{Favorskiy}},
  \bibinfo{author}{\bibfnamefont{D.}~\bibnamefont{Vu}},
  \bibinfo{author}{\bibfnamefont{E.}~\bibnamefont{Peytavit}},
  \bibinfo{author}{\bibfnamefont{S.}~\bibnamefont{Arscott}},
  \bibinfo{author}{\bibfnamefont{D.}~\bibnamefont{Paget}}, \bibnamefont{and}
  \bibinfo{author}{\bibfnamefont{A.~C.~H.} \bibnamefont{Rowe}},
  \bibinfo{journal}{Rev. Sci. Instr.} \textbf{\bibinfo{volume}{81}},
  \bibinfo{pages}{103902} (\bibinfo{year}{2010}).

\bibitem[{\citenamefont{Qi et~al.}(2006)\citenamefont{Qi, Yu, and
  Flatte}}]{flatte2006}
\bibinfo{author}{\bibfnamefont{Y.}~\bibnamefont{Qi}},
  \bibinfo{author}{\bibfnamefont{Z.~G.} \bibnamefont{Yu}}, \bibnamefont{and}
  \bibinfo{author}{\bibfnamefont{M.~E.} \bibnamefont{Flatte}},
  \bibinfo{journal}{Phys. Rev. Lett} \textbf{\bibinfo{volume}{96}},
  \bibinfo{pages}{026602} (\bibinfo{year}{2006}).

\bibitem[{\citenamefont{Finkel'shtein}(1983)}]{finkelshtein1983}
\bibinfo{author}{\bibfnamefont{A.~M.} \bibnamefont{Finkel'shtein}},
  \bibinfo{journal}{Sov. Phys. JETP} \textbf{\bibinfo{volume}{57}},
  \bibinfo{pages}{97} (\bibinfo{year}{1983}).

\bibitem[{\citenamefont{Sommerfeld}(1928)}]{sommerfeld1928}
\bibinfo{author}{\bibfnamefont{A.}~\bibnamefont{Sommerfeld}},
  \bibinfo{journal}{Z. Phys.} \textbf{\bibinfo{volume}{47}}, \bibinfo{pages}{1}
  (\bibinfo{year}{1928}).

\bibitem[{\citenamefont{Paget et~al.}(2012)\citenamefont{Paget, Cadiz, Rowe,
  Moreau, Arscott, and Peytavit}}]{paget2012}
\bibinfo{author}{\bibfnamefont{D.}~\bibnamefont{Paget}},
  \bibinfo{author}{\bibfnamefont{F.}~\bibnamefont{Cadiz}},
  \bibinfo{author}{\bibfnamefont{A.~C.~H.} \bibnamefont{Rowe}},
  \bibinfo{author}{\bibfnamefont{F.}~\bibnamefont{Moreau}},
  \bibinfo{author}{\bibfnamefont{S.}~\bibnamefont{Arscott}}, \bibnamefont{and}
  \bibinfo{author}{\bibfnamefont{E.}~\bibnamefont{Peytavit}},
  \bibinfo{journal}{Journal of Applied Physics} \textbf{\bibinfo{volume}{111}},
  \bibinfo{pages}{123720} (\bibinfo{year}{2012}).

\bibitem[{\citenamefont{Amo et~al.}(2007)\citenamefont{Amo, Vina, Lugli,
  Tejedor, Toropov, and Zhuravlev}}]{amo2007}
\bibinfo{author}{\bibfnamefont{A.}~\bibnamefont{Amo}},
  \bibinfo{author}{\bibfnamefont{L.}~\bibnamefont{Vina}},
  \bibinfo{author}{\bibfnamefont{P.}~\bibnamefont{Lugli}},
  \bibinfo{author}{\bibfnamefont{C.}~\bibnamefont{Tejedor}},
  \bibinfo{author}{\bibfnamefont{A.~I.} \bibnamefont{Toropov}},
  \bibnamefont{and} \bibinfo{author}{\bibfnamefont{K.~S.}
  \bibnamefont{Zhuravlev}}, \bibinfo{journal}{Phys. Rev. B}
  \textbf{\bibinfo{volume}{75}}, \bibinfo{pages}{085202}
  (\bibinfo{year}{2007}).

\bibitem[{\citenamefont{Zerrouati et~al.}(1988)\citenamefont{Zerrouati, Fabre,
  Bacquet, Bandet, Frandon, Lampel, and Paget}}]{zerrouati1988}
\bibinfo{author}{\bibfnamefont{K.}~\bibnamefont{Zerrouati}},
  \bibinfo{author}{\bibfnamefont{F.}~\bibnamefont{Fabre}},
  \bibinfo{author}{\bibfnamefont{G.}~\bibnamefont{Bacquet}},
  \bibinfo{author}{\bibfnamefont{J.}~\bibnamefont{Bandet}},
  \bibinfo{author}{\bibfnamefont{J.}~\bibnamefont{Frandon}},
  \bibinfo{author}{\bibfnamefont{G.}~\bibnamefont{Lampel}}, \bibnamefont{and}
  \bibinfo{author}{\bibfnamefont{D.}~\bibnamefont{Paget}},
  \bibinfo{journal}{Phys. Rev. B} \textbf{\bibinfo{volume}{37}},
  \bibinfo{pages}{1334} (\bibinfo{year}{1988}).

\end{thebibliography}

\end{document}